\documentclass[journal=eds]{CUP-JNL-DTM}%

\usepackage{natbib}

\usepackage[dvipsnames]{xcolor}
\usepackage{graphicx}
\usepackage{multicol,multirow}
\usepackage{amsmath,amssymb,amsfonts}
\usepackage{mathrsfs}
\usepackage{amsthm}
\usepackage{rotating}
\usepackage{appendix}
\usepackage{ifpdf}
\usepackage[T1]{fontenc}
\usepackage{newtxtext}
\usepackage{newtxmath}
\usepackage{textcomp}
\usepackage{xcolor}
\usepackage{lipsum}
\usepackage[colorlinks,allcolors=blue]{hyperref}
\usepackage{tabularx}
\usepackage{chngcntr}
\usepackage{lineno}

\theoremstyle{definition}

\numberwithin{equation}{section}

\jname{Full Paper Submission to Climate Informatics 2026}
\articletype{APPLICATION PAPER}
\jyear{2025}
\newcolumntype{B}{X} 
\newcolumntype{m}{>{\hsize=.75\hsize\raggedright\arraybackslash}X} 
\newcolumntype{s}{>{\hsize=.5\hsize\raggedright\arraybackslash}X} 
\newcolumntype{t}{>{\hsize=.25\hsize\raggedright\arraybackslash}X} 

\begin{document}

\begin{Frontmatter}

\title[Article Title]{Dissipating the correlation smokescreen: Causal decomposition of the radiative effects of biomass burning aerosols over the South-East Atlantic}

\author[1]{Emilie Fons}
\author[2,3,4]{Isabel L. McCoy}
\author[5]{Tom Beucler}
\author[1,6]{David Neubauer}
\author[1]{Ulrike Lohmann}

\authormark{Fons \textit{et al}.}

\address[1]{\orgdiv{Institute for Atmospheric and Climate Science}, \orgname{ETH Zurich}, \orgaddress{\city{Zurich},   \country{Switzerland}}}

\address[2]{\orgdiv{Cooperative Institute for Research in Environmental Sciences}, \orgname{University of Colorado}, \orgaddress{\city{Boulder},\state{Colorado},  \country{USA}}}

\address[3]{\orgdiv{Chemical Sciences Laboratory}, \orgname{NOAA}, \orgaddress{\city{Boulder},\state{Colorado},  \country{USA}}}

\address[4]{\orgdiv{Now at: Department of Atmospheric Science}, \orgname{Colorado State University}, \orgaddress{\city{Fort Collins},\state{Colorado},  \country{USA}}}

\address[5]{\orgdiv{Faculty of Geosciences and Environment}, \orgname{University of Lausanne}, \orgaddress{\city{Lausanne},   \country{Switzerland}}}

\address[6]{\orgdiv{Now at: Urban Climate}, \orgname{GeoSphere Austria}, \orgaddress{\city{Vienna},   \country{Austria}}}

\authormark{Fons et al.}



\abstract{Biomass burning aerosols (BBAs) from Southern Africa seasonally overlie the semi-permanent South-East Atlantic (SEA) stratocumulus deck, impacting the region's energy budget through complex aerosol-cloud-radiation-meteorology interactions. Climate model intercomparison initiatives, like the Aerosol Comparisons between Observations and Models (AeroCom), have highlighted the large inter-model variability for BBA radiative effects, especially over the SEA, due to parameterization of emission modeling and smoke properties. Observational constraints are needed to reduce these uncertainties, but correlative observational studies are typically affected by confounding meteorological influences. We propose a physically informed statistical approach, based on causal graphs applied to satellite observations, to disentangle BBA influences on shortwave radiation over the SEA and identify the main sources of statistical biases plaguing observational studies. We find that, during the fire season, BBAs cause a regional shortwave cooling of -2.5 W m$^{-2}$, which can be decomposed into equal contributions from three physical pathways: aerosol-radiation interactions (ARI), adjustments to ARI, and aerosol-cloud interactions (ACI). We also perform ablation experiments with graph variants to investigate the main sources of confounding --- like large-scale winds, humidity-biased retrievals or spatial aggregation of data --- and show that they result in biased radiative effect estimates (-50 \% and +15 \%). Once free of such biases, our derived causal estimates of smoke radiative effects can be used as observational constraints to improve climate models.}

\end{Frontmatter}

\section*{Impact Statement}
We apply a rigorous causal inference approach to identify and remove biases impeding the quantification of the radiative effect of smoke aerosols from earth observations over the South-East Atlantic, a region severely affected by biomass burning aerosols five months a year. Such a bias-free observational estimate can be used to constrain climate models struggling with the correct simulation of smoke-cloud-radiation interactions and of their impacts on climate change and precipitation extrema.

\section{Introduction}

Not only are biomass burning aerosols (BBAs) deleterious for air quality, they also modify the radiative budget of the Earth \citep{reidCriticalReviewHealth2016, szopa_2021_ipcc}. 
Smoke radiative effects obtained from climate models (e.g. AeroCom II models) suffer from uncertainties related to fire emission modeling and parameterization choices \citep[e.g.,][]{myhreRadiativeForcingDirect2013, zhongUsingModelledRelationships2022}. With increasing risks of wildfires due to global warming \citep{parmesan_2022_ipcc}, it is essential to constrain the uncertainties surrounding BBA radiative effects. 

\begin{figure}[t]
    \FIG{\includegraphics[width=0.5\textwidth]{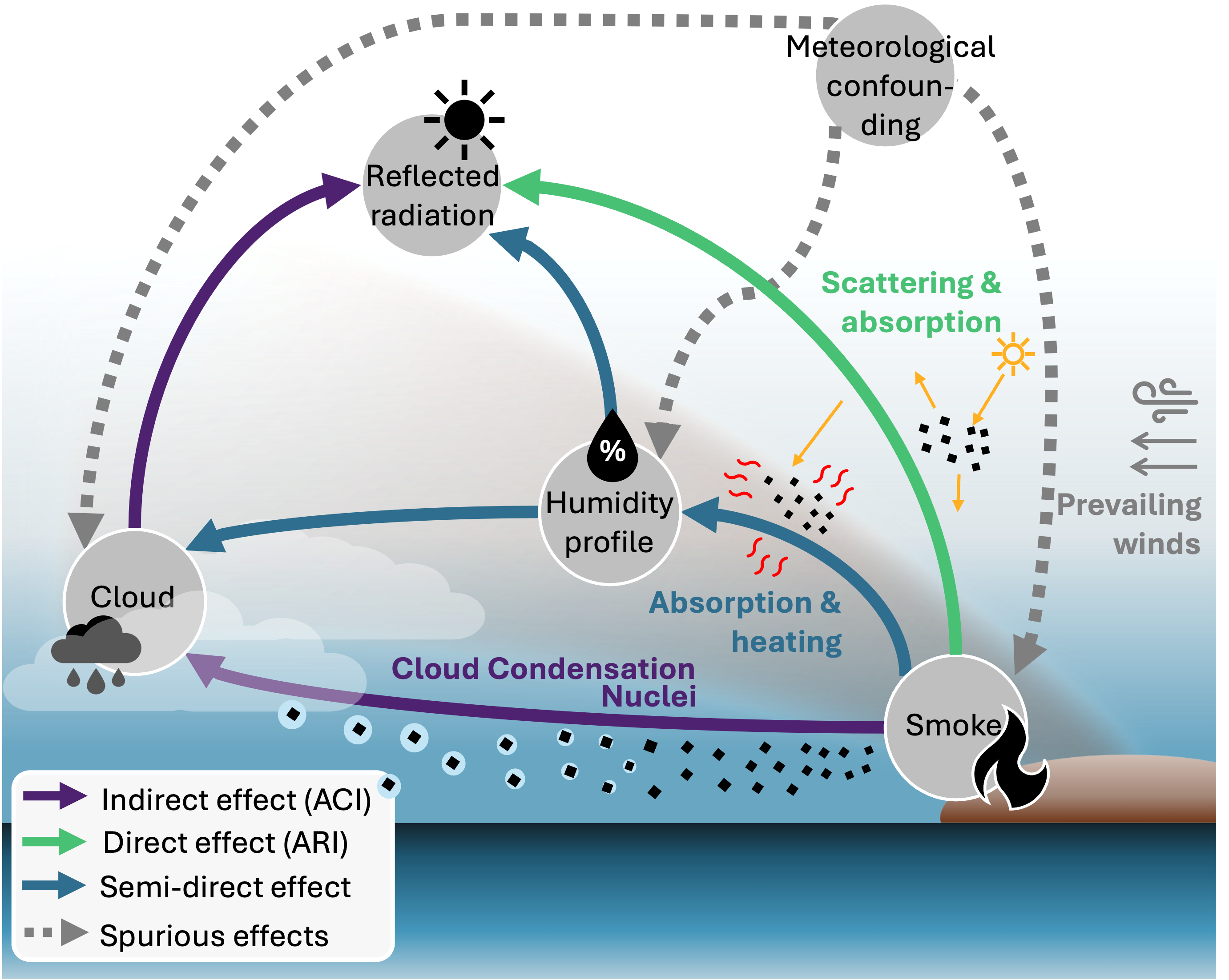}}
    \caption[Schematic causal pathways for the shortwave radiative effects of biomass burning aerosols]{Schematic causal pathways for the shortwave radiative effects of BBAs. The direct, indirect and semi-direct pathways are shown in green, purple and blue, respectively. In grey, dotted arrows show how confounding meteorological influences can bias the statistical estimate of BBA radiative effects}
    \label{fig:causal_graph_schematic}
\end{figure}

Due to the complexity of smoke-cloud-radiation-meteorology interactions, reducing this uncertainty is not trivial. One third of global BBA emissions originate from fires in Southern Africa during the biomass burning season \citep{vanmarleHistoricGlobalBiomass2017}. Large-scale winds blow the smoke from Southern Africa towards the marine boundary layer (BL) clouds of the South-East Atlantic (SEA). BBAs are composed mostly of Black Carbon (BC) and Organic Carbon (OC), with different radiative properties:  BC strongly absorbs shortwave (SW) radiation, OC mostly scatters SW \citep{reidReviewBiomassBurning2005}. Thus the sign of BBA radiative effects dependent on BBA composition, but also on the underlying surface: BBAs can have a SW cooling effect over dark surfaces like the ocean but a warming effect over bright surfaces like stratocumulus decks. Additionally, freshly emitted BC and OC are hydrophobic, but they can age during atmospheric transport, become hydrophilic and eventually serve as cloud condensation nuclei (CCN) for cloud formation \citep{reidReviewBiomassBurning2005}. Fig. \ref{fig:causal_graph_schematic} summarizes the  complexity of the three pathways by which BBAs interact with radiation: (1) the direct effect, also termed aerosol-radiation interactions (ARI), corresponding to the direct absorption and scattering of SW by aerosols; (2) the indirect effect, also termed aerosol-cloud interactions (ACI), describing how hygroscopic aerosols can serve as CCN and modify the radiative properties of clouds (rapid adjustment to ACI); (3) the semi-direct effect, considered a rapid adjustment to ARI, whereby aerosol absorption of SW radiation modifies the atmospheric profiles of temperature and humidity, thus impacting cloud formation \citep{bellouinBoundingGlobalAerosol2020}. Even though smoke can also interact directly with longwave (LW) radiation, the direct SW effect will dominate \citep[e.g.,][]{stone_empirical_2011}. Besides, SEA clouds will have a negligible LW effect at the top of the atmosphere due to their low altitude \citep{Lohmann_Lüönd_Mahrt_2016b}, hence the focus on SW radiation in this study. 

Climate models can be used to compute and decompose aerosol radiative effects as the difference between a control and a counterfactual scenario. Examples of counterfactual scenarios include switching off all anthropogenic aerosols emissions \citep{ghanTechnicalNoteEstimating2013} or switching off the ARI pathway by modifying a model's radiation or cloud scheme \citep{diamondCloudAdjustmentsLargescale2022}. However, individual parametrization choices lead to high inter-model variability of aerosol radiative forcing estimates, especially over the SEA \citep{stierHostModelUncertainties2013, myhreRadiativeForcingDirect2013}. Estimating radiative forcing from observations is not so straightforward, as co-variability of large-scale meteorological conditions induces confounding that biases the aerosol-cloud-radiation relationships (Fig. \ref{fig:causal_graph_schematic}). 

In this study, we use causal inference \citep{pearlCausalityModelsReasoning2009, imbensCausalInferenceStatistics2015} to compute BBA radiative forcing from 16 years of satellite observations and reanalysis data. We use the causal framework to rigorously justify the choice of meteorological control variables, allowing us to decompose the radiative forcing into its direct, indirect and semi-direct components, and to identify and estimate confounding biases originating from meteorological influences and methodological choices. 

\section{Data and Methods}

\subsection{Satellite and Reanalysis Data}
The dataset comprises 16 years (2003 - 2018) of cloud, aerosol and meteorology data for the South-East Atlantic (SEA) low-level cloud region, which is impacted by smoke during the biomass burning season (30$^{\circ}$W-15$^{\circ}$E- 25$^{\circ}$S-0$^{\circ}$, June-October).  

The cloud observations are obtained from the Moderate Resolution Imaging Spectroradiometer \citep[MODIS,][]{platnickMODISAtmosphereL32015} and the Clouds and the Earth’s Radiant Energy System \citep[CERES,][]{doelling_geostationary_2013, doelling_advances_2016} instrument aboard the Aqua satellite. The meteorological variables are downloaded from the ERA5 \citep{hersbachERA5HourlyData2018a, hersbachERA5HourlyData2018} and CAMS reanalyses \citep{inness_cams_2019} from ECMWF. The products used in this study are: daytime cloud fraction (CF), cloud optical depth (COD), cloud droplet effective radius ($r_\mathrm{eff}$) and Aerosol Optical Depth (AOD) from the daily level 3 $1^{\circ}\times1^{\circ}$ MODIS product; the reflected top-of-atmosphere (TOA) shortwave flux ($\mathrm{SW \uparrow}$) from the daily level 3 $1^{\circ}\times1^{\circ}$ CERES product; pressure-level wind speeds and relative humidity from ERA5; pressure-level mixing ratios of BC and OC from CAMS. Using the adiabaticity assumption \citep{brenguierRadiativePropertiesBoundary2000, quaasConstrainingTotalAerosol2006}, we derive the cloud depth $H$ from COD and $r_\mathrm{eff}$. 
The key variables are listed in Table \ref{tab:causal_nodes}. 

The data are co-located on the $1^{\circ}\times1^{\circ}$ grid of the MODIS data using a conservative regridding approach \citep{zhuang2023pangeo} and the reanalysis data are interpolated to the approximate local crossing time of the Aqua satellite. Among cloudy pixels, only low marine cloud pixels, identified as having a cloud top pressure larger than 680 hPa, are kept in the analysis. 

\begin{table}[htbp]
\tabcolsep=0pt%
\TBL{\caption[{Causal graph node variables}]{Causal graph node variables and their  descriptions}\label{tab:causal_nodes}}
{{\small
\begin{fntable}
\begin{tabularx}{1.0\textwidth}{c  X}
 \TCH{Graph node } & \TCH{ Description} \\ 
 \hline
 \TCH{BBA} & Biomass burning aerosols. By default, this is CAMS' total mass of BC and OC in the atmospheric column. Replaced by MODIS AOD to evaluate proxy confounding. \\  
 \TCH{$r_\mathrm{eff}$} & Cloud droplet effective radius from MODIS. \\
 \TCH{$H$} & Cloud depth, calculated from MODIS retrievals using the adiabaticity assumption. \\
 \TCH{$\mathrm{COD_w}$} & Cloud Optical Depth from MODIS, weighted by the grid-box cloud fraction (also from MODIS).\\ 
 \TCH{$\mathrm{SW\uparrow}$} & Reflected top-of-atmosphere SW radiation flux from CERES. \\
 \TCH{$\mathrm{RH_{BL}}$} & Relative humidity from ERA5, averaged over pressure levels below the boundary layer height, as identified using the eponymous variable in ERA5. \\ 
 \TCH{$\mathrm{U_{700}}$} & Eastward wind speed from ERA5, chosen at 700 hPa to capture mid-tropospheric large-scale transport by the Southern African Easterly Jet \citep[AEJ-S,][]{pistoneExploringElevatedWater2021, adebiyiRoleSouthernAfrican2016}.\\
\end{tabularx}
\end{fntable}}
}%
\end{table}

\subsection{Causal inference}
\begin{figure}[t]
    \FIG{\includegraphics[width=0.8\textwidth]{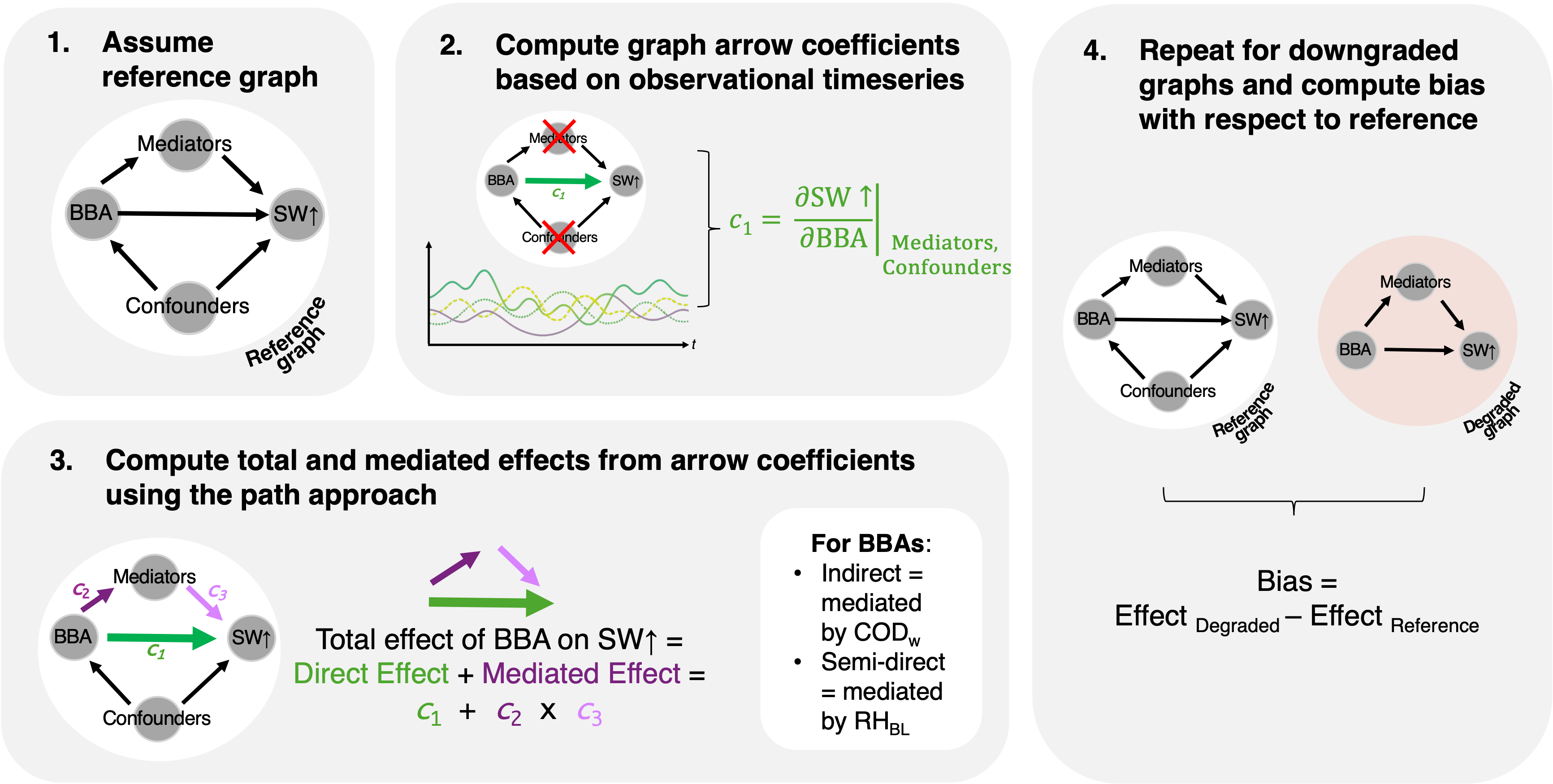}}
    \caption[Summary of the method]{Summary of the method, decomposed in 4 steps, from (1.) assumption of the reference causal graph to (4.) computation of biases in radiative effect estimates}
    \label{fig:method_summary}
\end{figure}

The causal workflow includes 4 steps, summarized in Fig. \ref{fig:method_summary}: 1) assumption of a groundtruth causal graph, 2) computation of partial sensitivities from observations, with control variables dictated by the graph, 3) computation of total and mediated effects from the partial sensitivities, and 4) comparison to other causal graphs and computation of resulting biases.  

Fig. \ref{fig:dce_agg} shows the `Reference graph', describing the pathways by which BBAs affect the SW radiation budget of the SEA region (Step 1). Its structure is hypothesized from expert knowledge and is not obtained from data-driven techniques such as causal discovery \citep{rungeCausalInferenceTime2023}. Like in Fig. \ref{fig:causal_graph_schematic}, the colors of the arrow identifiers $a_n$ highlight the direct (ARI), indirect (ACI, cloud-mediated) and semi-direct (adjustments to ARI, humidity-mediated) pathways through which BBAs modify SW radiation at TOA.

Causal effect estimation is done with the Python Tigramite package\footnote{\url{https://github.com/jakobrunge/tigramite}} following Wright's linear path method \citep{wrightCorrelationCausation1921, rungeIdentifyingCausalGateways2015}. The estimation starts with the derivation of partial sensitivities, i.e., the values for each causal arrow in the graph, yielding the arrow colors in Fig. \ref{fig:dce_agg} (Step 2). Once the causal links have been estimated, causal effects between two variables that are not directly linked by an arrow are calculated by tracing the path(s) that link the two variables and multiplying the coefficients along these paths (Step 3). This is formalized in Eq. \ref{eq:sw_forcing2}, which shows how the total shortwave radiative effect of BBAs is obtained with a linear multiplication of partial senstivities: 

{\small
\begin{align}
\mathrm{RE_{BBA\rightarrow SW}} =
- \mathrm{\frac{dSW\uparrow}{dBBA}}\times\Delta \mathrm{BBA_{Forcing},\, where: } 
\underbrace{\mathrm{\frac{dSW\uparrow}{dBBA}}}_{\mathrm{Total\,effect}}
= \color{Green}{\mathrm{\left.\frac{\partial SW\uparrow}{\partial BBA}\right\rvert_{direct}}}
+ \color{Purple}{\mathrm{\left.\frac{\partial SW\uparrow}{\partial BBA}\right\rvert_{indirect}}}
+ \color{MidnightBlue}{\mathrm{\left.\frac{\partial SW\uparrow}{\partial BBA}\right\rvert_{semi-direct}}}
\nonumber \\
= \mathrm{
\color{Green}{\frac{\partial SW\uparrow}{\partial BBA}}
+ \color{Purple}{\frac{\partial SW\uparrow}{\partial COD_w}\frac{\partial COD_w}{\partial BBA}}
+ \color{MidnightBlue}{\frac{\partial SW\uparrow}{\partial RH_{BL}}\frac{\partial RH_{BL}}{\partial BBA}}
}
= \color{Green}{a_1}
+ \color{Purple}{a_2 a_3 a_4}
+ \color{MidnightBlue}{\underbrace{a_5 a_6}_{\substack{\mathrm{non\text{-}cloud}\\\mathrm{semi\text{-}direct}}}}
+ \underbrace{
\color{MidnightBlue}{
\left(
a_5 a_7
+ a_5 a_8
\left(
a_9 + a_{10}\color{Purple}{a_3}\color{MidnightBlue}{}
\right)
\right)
}
\color{Purple}{a_4}
}_{\substack{\mathrm{cloud\text{-}mediated}\\\mathrm{semi\text{-}direct}}}
\color{black}{,}
\nonumber \\
\text{and:\,}
\Delta\mathrm{BBA_{Forcing}}
= \overline{\text{BBA}_{\substack{25\%\,\text{Smokiest}\\\text{days}}}}
- \overline{\text{BBA}_{\substack{25\%\,\text{Least}\\\text{smoky\,days}}}}
\,(\mathrm{see\,Fig.}\ \ref{fig:cloud_vs_smoke})
\label{eq:sw_forcing2}
\end{align}
}

To estimate the impact of the main sources of statistical biases (Step 4), we  repeat Steps 1 to 3 for `degraded' variants of the graph, as described in Table \ref{tab:causal_graphs} and Fig. \ref{fig:all_causal_graphs}.

\begin{table}[htbp]
\TBL{\caption[Description of the causal graphs considered in this study]{Description of the causal graph variants considered in this study}
\label{tab:causal_graphs}}
{
{\small
\begin{fntable}
\begin{tabularx}{1.0\linewidth}{s m X}
 Graph name & Arrows in Fig. \ref{fig:dce_agg} & Description \\ 
 \hline
 Reference graph & All arrows in Fig. \ref{fig:dce_agg} & Reference graph against which all the other confounded graphs are evaluated \\  
 AOD confounding & Same as reference graph, only the data source changes & AOD is used as a proxy for the BBA variable, introducing confounding as AOD is dependent upon $\mathrm{RH_{BL}}$ \citep{stierLimitationsPassiveRemote2016,gryspeerdtConstrainingAerosolInfluence2016} \\
 Wind confounding & Reference graph minus the arrows pointing out of $\mathrm{U_{700}}$ & The westward AEJ-S, approximated by $\mathrm{U_{700}}$, transports smoke, but also humidity, and influences subsidence and boundary layer circulations \citep{adebiyiRoleSouthernAfrican2016, pistoneExploringElevatedWater2021}. Omitting $\mathrm{U_{700}}$ thus introduces confounding.  \\
 Regime confounding &  Reference graph, but $r_\mathrm{eff}$ and $H$ are omitted and  replaced by a direct arrow from BBA to $\mathrm{COD_w}$ & Regimes of clouds with different depths are not explicitly separated, causing confounding \citep{fonsStratocumulusAdjustmentsAerosol2023}.  \\ 
 Auto-dependency & Reference graph minus the autodependency links & For all variables X, \textit{X}(\textit{t} + 1 day) is considered independent from \textit{X}(\textit{t}) \\
 Lagged feedbacks & Reference graph minus the lagged  links from $\mathrm{COD_w}$ to BBA and $\mathrm{RH_{BL}}$ & Lagged effects of clouds on aerosol and humidity fields are removed \\
 Spatial aggregation & Same as `Reference graph', but the data normalization and the graph coefficients are calculated using the `multidata' Tigramite option and not separately in each $3^{\circ}\times3^{\circ}$ gridbox & The results are spatially generalizable but do not take into account potential latent spatially varying meteorological confounders \\ 
 Spatial aggregation + Longitude & Same as `Spatial Aggregation' but  `Longitude' is added as a graph node which confounds all the other variables & A very simple form of Earth Embedding aimed at reducing confounding by spatial aggregation. This acknowledges that different meteorogology-cloud-aerosol regimes exist, mostly along the longitude axis.  

\end{tabularx}
\end{fntable}}
}%
\end{table}

\subsection{Practical considerations: linearity, spatial aggregation, uncertainty intervals}

It can be questioned whether smoke-cloud-radiation interactions are linear \citep[e.g., see][]{dohertyModeledObservedProperties2022}. \cite{lohmannCanDirectSemidirect2001} found that linear additivity of the direct, indirect and semi-direct radiative effects of BBAs was a valid assumption on a global scale. To test this with our regional data, we do a multiple linear regression of $\mathrm{SW\uparrow}$ with respect to its parents variables in the reference graph, namely, BBA, $\mathrm{COD_w}$, $\mathrm{RH_{BL}}$ and $\mathrm{SW\uparrow}(t-1)$. On average for the region, the parent variables explain 76 \% of the variability in the TOA SW. 
We thus consider the linear approximation to be reasonable. 

Because aggregation over large regions can induce spurious statistical associations due to spatially covarying meteorological influences \citep{grandeyCriticalLookSpatial2010}, we do not carry out the causal effect estimation of the large SEA region at once, but rather in sub-domains \citep{loebObservationalStudyRelationship2008} of size $3^{\circ}\times3^{\circ}$, each containing 9 ($1^{\circ}\times1^{\circ}$) grid points. The data are normalized and de-seasonalized by removing the mean seasonal cycle and dividing by the standard deviation of the seasonal cycle in each sub-domain. This means that the structure of the causal graph is spatially generalizable over the SEA region, but the causal effects are not. 
With the `Spatial Aggregation' graph of Table \ref{tab:causal_graphs}, we bypass this spatial segmentation and pool all sub-domains using multidata causal inference to compute spatially uniform causal effects. `Spatial Aggregation + Longitude' is an attempt at correcting the resulting aggregation bias while retaining spatial generalization of the results.

 The uncertainties for the radiative effect estimates are computed with standard error propagation of the 68.2 \% ($\pm\sigma_\mathrm{dev}$) bootstrap error intervals (with 100 bootstrap members) around the estimates of the causal links. Maps showing causal effects and radiative effects feature stippling where the effects are not statistically different from 0, i.e., where the confidence interval includes 0.

\section{Results and discussion}
\subsection{Causal graph validation}

\begin{figure}[htbp]
    \FIG{\includegraphics[width=0.7\textwidth]{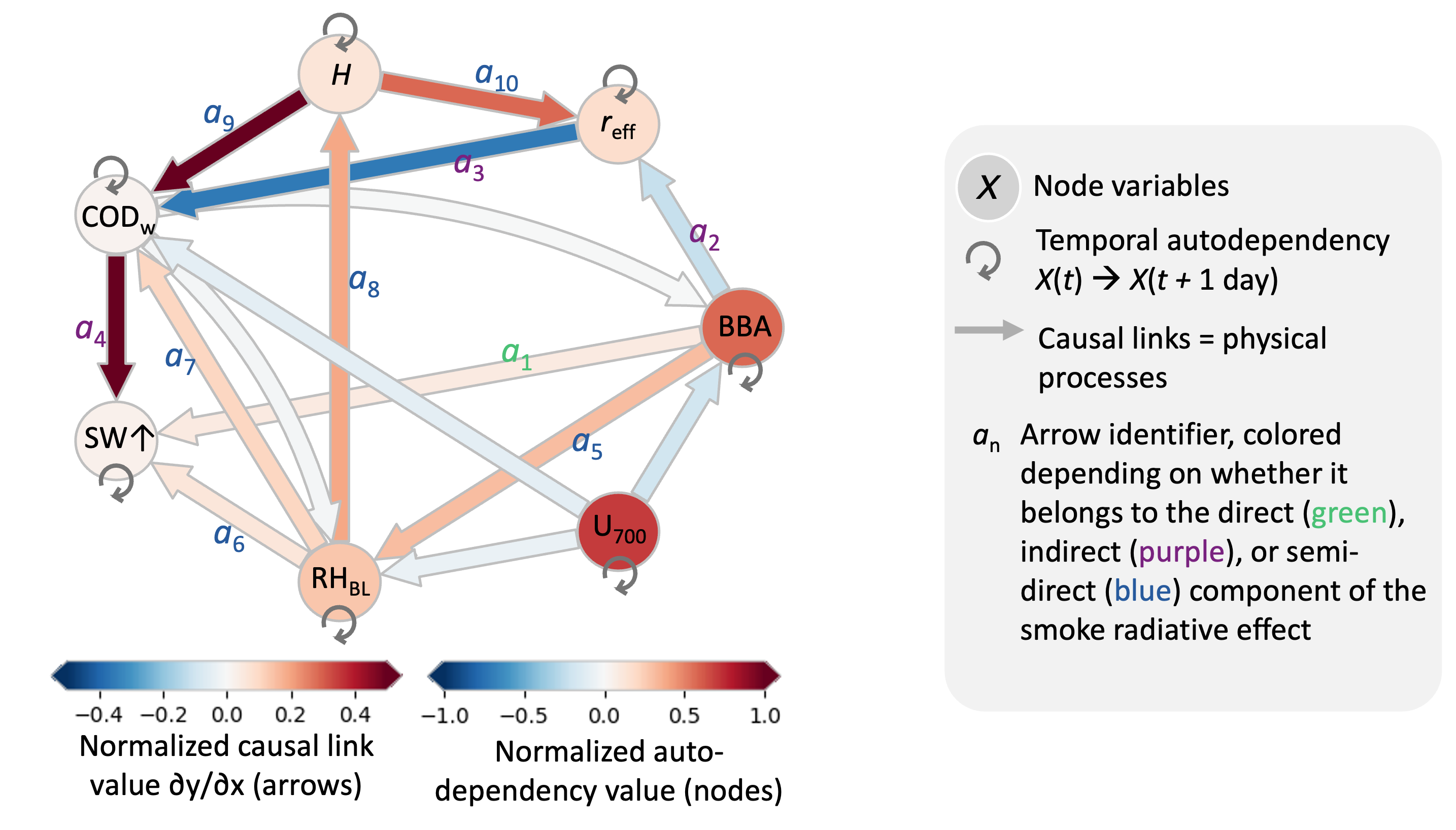}}
    \caption[Quantified causal links for the BBA-SW$\uparrow$ reference causal graph]{Quantified causal links for the reference graph. Arrow colors correspond to the  normalized magnitude of causal links between two variables, for contemporaneous effects (straight arrows) and lagged effects (curved arrows), while node colors correspond to magnitude of autodependency links}
    \label{fig:dce_agg}
\end{figure}

Fig. \ref{fig:dce_agg} shows the computed values of every causal arrow featured in the reference causal graph (Step 2 in Fig. \ref{fig:method_summary}). The causal links are resolved spatially on a $3^{\circ}\times3^{\circ}$ grid (Fig. \ref{fig:dce_maps}) but a domain-wide mean is shown here. The analysis of these values constitutes a physical validation of the method as the signs of the effects correspond to the direction of the expected physical mechanisms (Table \ref{tab:dce_interpretations}). 

\begin{table}[]
\TBL{\caption[Description of the graph's causal links and the physical interpretation of their sign]{Physical interpretation of the reference graph's causal links (Fig. \ref{fig:dce_agg})}
\label{tab:dce_interpretations}}
{{\small
\begin{fntable}
\begin{tabularx}{1.0\textwidth}{ t c B }
 Arrow in the & Computed & Physical Interpretation \\
 Reference Graph & Sign & \\
 \hline
 BBA $\rightarrow$  SW$\uparrow$  & + & Direct aerosol effect (some absorption by BC in the east of the domain but mostly scattering by OC in the west) \\
  $\mathrm{COD_w} \rightarrow$  SW$\uparrow$  & + & Strong albedo of clouds over ocean \\
  $\mathrm{RH_{BL}}$ $\rightarrow$  SW$\uparrow$  & + & The only effect that is not as expected. Could be due to a confounding influence from temperature \citep[e.g., less efficient water vapor absorption of SW at colder temperature,] []{wangTemperaturedependenceNearUVAbsorption2022}, or an imperfect separability of water vapor and cloud effects due to method limitations. \\

 BBA $\rightarrow$  $r_\mathrm{eff}$ & - & Aged hygroscopic BBAs can serve as CCN, leading to droplet size reduction at a constant liquid water content \citep{guptaImpactVariabilityVertical2021} \\
   BBA $\rightarrow$  $\mathrm{RH_{BL}}$ & + & When above the BL, BBA can cause increased LTS, less entrainment and a moister BL \citep{kochBlackCarbonSemidirect2010, wilcoxStratocumulusCloudThickening2010} \\

 H  $\rightarrow$  $r_\mathrm{eff}$  & + & Adiabatic cloud assumption for thin low-level clouds \citep{Lohmann_Lüönd_Mahrt_2016} \\
 $\mathrm{RH_{BL} \rightarrow H}$ & + & Cloud deepening in moist BL \\
  $r_\mathrm{eff}$ $\rightarrow \mathrm{COD_w}$  & - & Twomey effect and potential positive cloud fraction adjustments \citep{twomeyInfluencePollutionShortwave1977, gryspeerdtConstrainingAerosolInfluence2016} \\
  H $\rightarrow  \mathrm{COD_w}$  & - & Depth-dependence of COD (by definition) \\
  $\mathrm{RH_{BL}}$ $\rightarrow \mathrm{COD_w}$  & + & RH-dependence of cloud formation \citep{kohler1921kondensation} \\
 $\mathrm{U_{700} \rightarrow BBA}$ & - & Westward smoke transport by the AEJ-S \citep{adebiyiRoleSouthernAfrican2016} \\ 
  $\mathrm{U_{700}}$ $\rightarrow \mathrm{COD_w}$  & - & Secondary circulations induced by the AEJ-S \citep{adebiyiRoleSouthernAfrican2016} \\
  $\mathrm{U_{700}}$ $\rightarrow$  $\mathrm{RH_{BL}}$ & - & Humidity advection \citep{pistoneExploringElevatedWater2021, baroperez2024ComparingInluenceofBBA} and secondary circulations induced by the AEJ-S \citep{adebiyiRoleSouthernAfrican2016} \\
    $\mathrm{COD_w} \rightarrow$ BBA (lag 1) & - & Wet scavenging of below-cloud smoke by precipitation \\  
    $\mathrm{COD_w} \rightarrow$ $\mathrm{RH_{BL}}$ (lag 1) & +/- & A superposition of various effects: precipitation, surface cooling, and turbulent mixing induced by cloud top cooling \\

  \end{tabularx}
  \end{fntable}}
  }%
\end{table}

\subsection{Total shortwave effect of biomass burning aerosols}

\begin{figure}[htbp]
    \FIG{\includegraphics[width=1.0\linewidth]{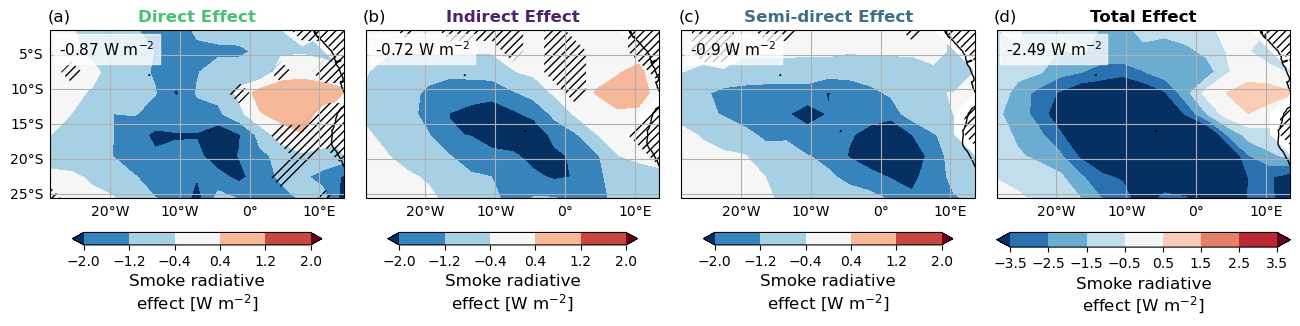}
    }
    \caption[BBA radiative effects per pathway: direct, indirect, semi-direct and resulting total]{BBA radiative effect per pathway: \textbf{(a)} direct, \textbf{(b)} indirect, \textbf{(c)} semi-direct and \textbf{(d)} resulting total (note the different colorscale). The region-average BBA shortwave radiative effect is displayed in the upper left corner. Stippling indicates where the bootstrap confidence intervals include 0} 
    \label{fig:total_effects}
\end{figure}

Fig. \ref{fig:total_effects} shows the total SW radiative effect of BBAs, as well as the fractions of that effect that are direct (non-mediated), indirect (cloud-mediated), and semi-direct ($\mathrm{RH_{BL}}$-mediated), as calculated using the path approach applied to the causal links computed just above (Step 3 in Fig. \ref{fig:method_summary}). 
The smoke emitted over the SEA stratocumulus region during the biomass burning season causes a net regional cooling of $\approx -2.5\,\mathrm{W\,m^{-2}}$. The direct, indirect and semi-direct pathways contribute almost equally to this cooling.

The warming/cooling dipole in total BBA radiative effect from east to west is mostly attributable to the direct radiative effect. This is due to the presence of dark absorbing aerosols overlying a dense bright stratocumulus deck in the east, leading to a local warming of up to $\mathrm{\approx 1\,W\,m^{-2}}$. In the west, cloud cover is sparser, and organic aerosols have photochemically `whitened' and become more efficient at scattering during atmospheric transport \citep[e.g.,][]{carterInvestigatingCarbonaceousAerosol2021}, leading to local cooling effects of more than $\mathrm{- 2\,W m^{-2}}$. Similar patterns have been found in reanalysis and model data, but model results are highly dependent on their choice of physical parameters and forcing methodologies \citep{myhreRadiativeForcingDirect2013, zuidemaSmokeCloudsSoutheast2016, malletDirectSemidirectRadiative2020, giuffrida2025BBAradiativeeffects}. Other observational studies \citep{kacenelenbogenEstimationsGlobalShortwave2019, dohertyModeledObservedProperties2022} have found stronger positive direct radiative effects of BBAs over the eastern SEA. However, the sampling of high-smoke events in these studies could have resulted in selection bias, whereas our estimates are representative of the whole duration of the biomass burning season in the SEA. 

The indirect radiative effect is negative in the western part of the domain but slightly positive at the coast. In the west, this is due to aged and hygroscopic smoke being entrained into the boundary layer and acting as CCN \citep{guptaImpactVariabilityVertical2021}, making clouds brighter and triggering cloud adjustments, like increased cloud cover and lifetime. In the east, freshly emitted BBAs are hydrophobic and cannot act as CCN \citep{redemannOverviewORACLESObseRvations2021}. If hydrophobic smoke is entrained into the boundary layer, it competes with hydrophylic aerosols for water vapor, resulting in fewer and less bright cloud droplets \citep{ching_black_2016}. However, it is also possible that thick smoke layers overlying clouds bias the retrievals of cloud properties \citep{alfaro-contrerasEvaluatingImpactAerosol2014}. Modeling studies have also found negative indirect smoke radiative effects of the same magnitude, although some studies have a simplistic parameterization of aerosol-cloud interactions \citep{rapNaturalAerosolDirect2013, lohmannIndirectEffectSulfate2000, giuffrida2025BBAradiativeeffects}.

While models have been shown to disagree on the sign and magnitude of the semi-direct effect \citep{malletDirectSemidirectRadiative2020, allenObservationallyConstrainedAerosol2019, sakaedaDirectSemidirectAerosol2011}, we detect uniform cooling, primarily driven by the cloud-mediated semi-direct effect (-0.7 W m$^{-2}$) with a smaller contribution from the non-cloud semi-direct effect (-0.2 W m$^{-2}$). 

Our observational results highlight that the total smoke radiative effect results from almost equally cooling contributions of the direct, indirect and semi-direct pathways. This finding can be used to constrain climate models that have previously found that the indirect effect dominates the total cooling \citep[e.g.,][]{lohmannCanDirectSemidirect2001, luBiomassSmokeSouthern2018}.

\subsection{Sources of confounding}

\begin{figure}[t]
    \FIG{\includegraphics[width=0.55\linewidth]{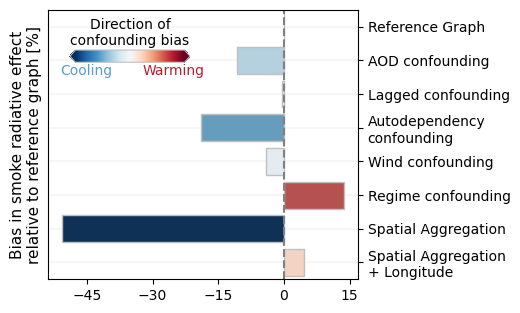}
    }
    \caption[Biases in smoke radiative effect caused by different sources of confounding]{Biases in smoke radiative effect caused by different sources of confounding}
    \label{fig:biases_bar}
\end{figure}

In Step 4 of the analysis, the reference graph is modified (Table \ref{tab:causal_graphs}) to single out sources of confounding bias. 

Spatial aggregation results in the largest confounding bias with a -51 \% cooling bias compared to the reference graph. This is mostly due to the longitudinal gradient of meteorological regimes present in the SEA, with very bright clouds and a thick smoke layer in the east, but more dispersed clouds and less aerosols in the pristine western part of the domain, creating a spurious positive correlation between aerosols, cloud brightness and reflected radiation. However, when simply adding longitude as a confounder node in the causal graph, the large bias almost vanishes (from -51\% to +5\%). This simple yet powerful change allows for spatial generalization while keeping bias at a minimum. 

\cite{diamondTimedependentEntrainmentSmoke2018} pointed out the time-dependence of smoke-cloud interactions over the SEA. Here we detect that neglecting auto-dependency of meteorological fields, and especially of $\mathrm{RH_{BL}}$, $\mathrm{U_{700}}$ and BBA (see node colors in Fig. \ref{fig:dce_agg}) leads to a 19\% overestimation of the cooling influence of BBAs. It would be interesting to repeat the present analyses with a Lagrangian perspective to identify the exact processes responsible for this behavior. 

Omitting $r_\mathrm{eff}$ as a mediator of the indirect radiative effect of BBAs leads to an underestimation of the cooling influence of BBAs (by 14\%). This is consistent with the findings of \cite{fonsStratocumulusAdjustmentsAerosol2023}, and is due to an overestimation of cloud optical depth decreases when aggregating data over different cloud regimes. 

The AOD cooling bias (-11\%) is due to the humidity-dependence of the AOD proxy, resulting in a spuriously enhanced correlation between RH and AOD \citep[e.g.,][]{stierLimitationsPassiveRemote2016, gryspeerdtConstrainingAerosolInfluence2016}.

Neglecting confounding from large-scale winds leads to a small overestimation of the cooling influence of BBAs. 
This is coherent with \cite{baroperez2024ComparingInluenceofBBA}, who found a radiative bias due to moisture accompanying the smoke plume, although their cooling bias was much stronger than ours.  

Importantly, while different sources of confounding might compensate each other to some degree, ignoring any of them would result in a loss of robustness as there is a lack of generalization power when confounding is present.  

\label{LastPage}
\section{Conclusions and perspectives}

We applied causal inference to estimate the observed radiative effect of smoke aerosols over the South-East Atlantic. We find that BBAs cause a total radiative effect of -2.5 W m$^{-2}$ with approximately equal contributions from the direct, indirect and semi-direct effects. Contrary to many modeling studies that find strong warming direct effects of BBAs, we find moderate warming (up to 1 W m$^{-2}$) that is confined in the east of the SEA basin, where freshly emitted smoke overlies a bright stratocumulus decks. Using variants of our causal graph, we identify the main sources of confounding: aggregation across space and cloud regimes, temporal autocorrelation, using AOD as a proxy for BBA aerosols and large-scale dynamical influences from the Southern-African Easterly Jet. This graph comparison highlights the key physical variables that need to be taken into account for smoke-cloud-radiation studies and for which measurement errors need to be reduced in future satellite developments or field campaigns. Some limitations of this study deserve to be investigated in future work, for instance, the ground truth assumption for the reference causal graph, the linear assumption, the Eulerian nature of the analyses, or satellite retrieval biases. Despite these limitations, causal inference proves to be a helpful tool to decompose the complex pathways through which biomass burning aerosols modify the radiative budget of the South-East Atlantic under complex confounding influences from meteorology, providing bias-free observational constraints that can guide process-based evaluations of climate models. Climate models with better smoke parameterizations will eventually be useful to evaluate the effects of wildfire smoke and pollution in the anthropogenic era. At the global scale, this could help increase confidence in future temperature projections. At the regional scale, this could help evaluate smoke- and pollution-driven changes in the hydrological cycle and thus inform adaptation measures related to droughts or flooding. \\

\begin{Backmatter}

\newpage

\paragraph{Funding Statement}
This work was supported by the European Union’s Horizon 2020 research and innovation program under Marie Sklodowska-Curie grant agreement No. 860100 (iMIRACLI). ILM research was supported by NOAA cooperative agreement NA22OAR4320151, for the Cooperative Institute for Earth System Research and Data Science (CIESRDS). We acknowledge the International Space Science Institute (ISSI) in Bern, Switzerland who facilitated discussions of these concepts through ISSI International Team Project $\#$23-576. 

\paragraph{Competing Interests}
The authors declare no competing interest.

\paragraph{Code and Data Availability Statement}
The source data used in this study can be downloaded from: MODIS (\url{https://ladsweb.modaps.eosdis.nasa.gov/archive/allData/61/MOD08_D3/}), CERES (\url{https://ceres.larc.nasa.gov/data/}), ERA5 (\url{https://cds.climate.copernicus.eu/datasets/reanalysis-era5-pressure-levels}), CAMS (\url{https://ads.atmosphere.copernicus.eu/datasets/cams-global-reanalysis-eac4}). The analysis and plotting code is available on Zenodo \citep{fons_software_2026}. Intermediate data produced by and used by the code are available on Zenodo as well \citep{fons_data_2026}.      

\paragraph{Ethical Standards}
The research meets all ethical guidelines, including adherence to the legal requirements of the study country.

\paragraph{Author Contributions}
Conceptualization: E.F.; U.L.; D.N. Methodology: E.F.; I.L.M.; T.B. Data curation: E.F. Data visualisation: E.F.. Writing original draft: E.F. All authors approved the final submitted draft. 

\newpage
\begin{appendix}
\section{Appendix}\label{appendixA}
\counterwithin{figure}{section}
\setcounter{figure}{0}    

\begin{figure}[htbp]
    \FIG{\includegraphics[width=0.9\textwidth]{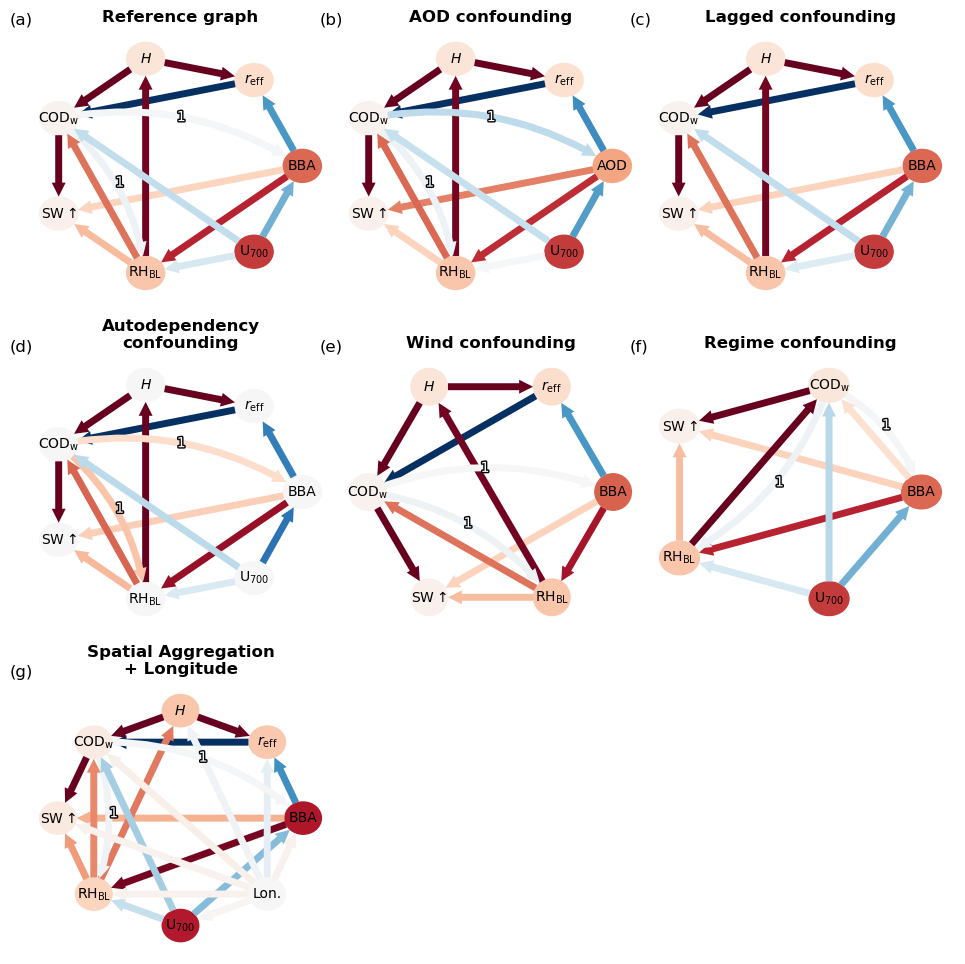}}
    \caption[]{All causal graph variants evaluated in the study, as described in Table \ref{tab:causal_graphs}. The arrow colorscale is bounded by -0.2 and 0.2, the node colorscale by -1 and 1}
    \label{fig:all_causal_graphs}
\end{figure}

\begin{figure}[htbp]
    \FIG{\includegraphics[width=0.75\textwidth]{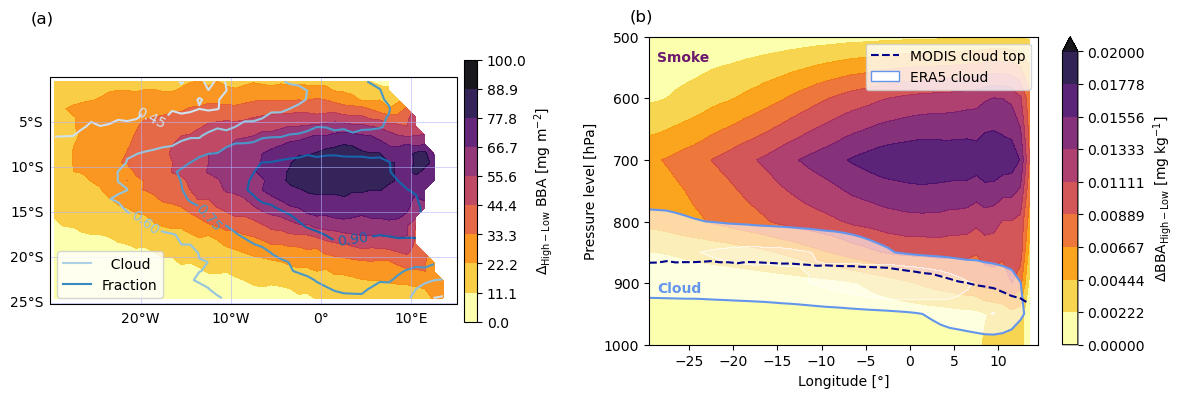}}
    \caption[]{Smoke and cloud fields over the SEA during the biomass burning season. (a) Filled contours show $\Delta\mathrm{BBA_{Forcing}}$ (defined in Eq. \ref{eq:sw_forcing2}) with overlain contours of MODIS cloud fraction. (b) Same but as a vertical cross-section, averaged
over latitude. Overlain are the mean low-level cloud layer computed from ERA5 liquid water content ($\geq 10 \mathrm{g\,kg^{-1}}$), and the mean cloud top pressure from MODIS}
    \label{fig:cloud_vs_smoke}
\end{figure}

\begin{figure}[htbp]
    \FIG{\includegraphics[width=0.65\textwidth]{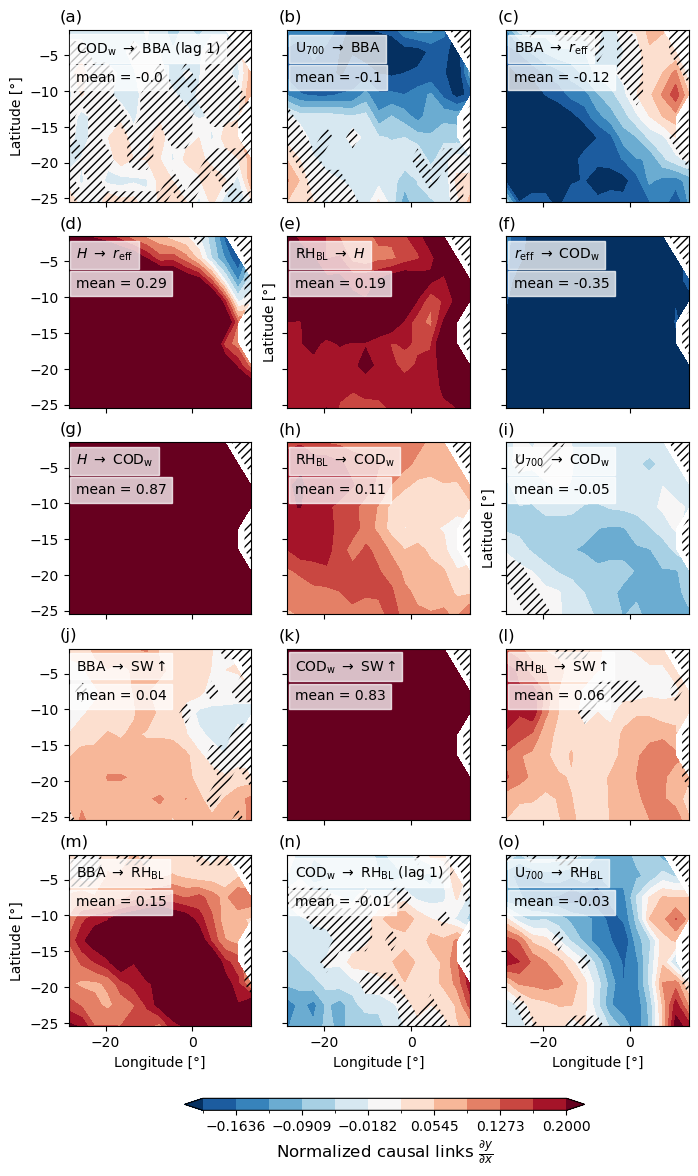}}
    \caption[]{Spatial distribution of normalized causal links for the reference graph. The subplots show the sign and magnitude of causal effects over the SEA region, for each causal arrows in the graph (Fig. \ref{fig:dce_agg}), as indicated in the upper left corner of each subplot}
    \label{fig:dce_maps}
\end{figure}

\end{appendix}

\newpage
\bibliographystyle{plainnat}  
\bibliography{references}
\end{Backmatter}

\end{document}